\documentclass{elsart4-1}

\usepackage{graphicx}
\usepackage{epstopdf}
\usepackage{amssymb}
\usepackage{amsmath}

\usepackage[english,francais]{babel}

\newtheorem{e-proposition}[theorem]{Proposition}

\newtheorem{e-definition}[theorem]{Definition\rm}

\renewcommand{\theequation}{\arabic{equation}}
\newcommand{\bea}{\begin{eqnarray}}
\newcommand{\eea}{\end{eqnarray}}
\newcommand{\be}{\begin{equation}}
\newcommand{\ee}{\end{equation}}

\newcommand{\rmL}{{\rm L}}
\newcommand{\rmR}{{\rm R}}

\newcommand{\rmH}{{\rm H}}

\newcommand{\kT}{k_{\rm B}T}

\def\og{\leavevmode\raise.3ex\hbox{$\scriptscriptstyle\langle\!\langle$~}}
\def\fg{\leavevmode\raise.3ex\hbox{~$\!\scriptscriptstyle\,\rangle\!\rangle$}}

\begin{document}


\centerline{to appear in Comptes Rendue Physique}
\begin{frontmatter}


\selectlanguage{english}

\title{Thermoelectrics with Coulomb coupled Quantum Dots}


\selectlanguage{english}
\author[authorlabel1]{Holger Thierschmann},
\ead{h.r.thierschmann@tudelft.nl}
\author[authorlabel2]{Rafael S\'anchez},
\ead{rafsanch@ing.uc3m.es}
\author[authorlabel3]{Bj\"orn Sothmann}
\ead{bjoern.sothmann@uni-wuerzburg.de}
\author[authorlabel4]{Hartmut Buhmann},
\ead{hartmut.buhmann@physik.uni-wuerzburg.de}
\author[authorlabel4]{Laurens W. Molenkamp}
\ead{laurens.molenkamp@physik.uni-wuerzburg.de}

\address[authorlabel1]{Kavli Institut of Nanoscience, Faculty of Applied Sciences, Delft University of Technology, Lorentzweg 1, 2628 CJ Delft The Netherlands}
\address[authorlabel2]{Instituto Gregorio Mill\'an, Universidad Carlos III de Madrid, 28911 Legan\'es, Madrid, Spain}
\address[authorlabel3]{Insitute for Theoretical Physics and Astrophysics, University of W\"urzburg, Am Hubland, 97074 W\"urzburg, Germany}
\address[authorlabel4]{Experimentelle Physik 3, Physikalisches Institut, Universit\"at W\"urzburg, Am Hubland, 97074 W\"urzburg, Germany}

\medskip
\begin{center}
{\small }
\end{center}

\begin{abstract}
In this article we review the thermoelectric properties of three terminal devices with Coulomb coupled quantum dots (QDs) as observed in recent experiments \cite{Thierschmann_Thermal_2015,Thierschmann_ThreeTerminal_2015}. The system we consider consists of two Coulomb-blockade QDs one of which can exchange electrons with only a single reservoir (heat reservoir) while the other dot is tunnel coupled to two reservoirs at a lower temperature (conductor). The heat reservoir and the conductor interact only via the Coulomb-coupling of the quantum dots. It has been found that two regimes have to be considered. In the first one heat flow between the two systems is small. In this regime thermally driven occupation fluctuations of the hot QD modify the transport properties of the conductor system. This leads to an effect called \textit{thermal gating}. Experiment have shown how this can be used to control charge flow in the conductor by means of temperature in a remote reservoir. We further substantiate the observations with model calculations and implications for the realization of an all-thermal transistor are discussed. In the second regime, heat flow between the two systems is relevant. Here the system works as a nano scale heat engine, as proposed recently \cite{Sanchez_Optimal_2011}. We review the conceptual idea, its experimental realization and the novel features arising in this new kind of thermoelectric device such as decoupling of heat and charge flow.    


\vskip 0.5\baselineskip

\selectlanguage{francais}
\vskip 0.5\baselineskip
\noindent
{\bf }

\end{abstract}
\end{frontmatter}


\selectlanguage{english}

\section{Introduction}

The search for new and more efficient ways of controlling heat flow and harvesting thermal energy was newly triggered in the past decades by the advances in fabrication of nano-devices. It has developed into an exciting field in solid state research~\cite{White_Beyond_2008,Shakouri_Recent_2011}. One promising route was found to lie in the strong thermoelectric response of mesoscopic devices due to their large Seebeck coefficients~\cite{Mahan_Best_1996,Radousky_Energy_2012}. This route has been explored fruitfully in recent years. Detailed experimental tests have provided new insight into thermoelectrics of small structures such as quantum point contacts and wires \cite{Molenkamp_Quantum_1990,Molenkamp_Peltier_1992,Riha_Heat_2016}, quantum dots (QDs) \cite{Staring_Coulomb_1993,Dzurak_Observation_1993,Dzurak_Thermoelectric_1997,Godijn_Thermopower_1999,Scheibner_Kondo_2005,Scheibner_Sequential_2007,Svensson_Lineshape_2012,Svensson_Nonlinear_2013} and double quantum dots \cite{Thierschmann_Diffusion_2013}. Numerous proposals indicate that strongly enhanced thermoelectric efficiencies can be achieved by using the properties of nanoscale conductors \cite{Hicks_Effect_1993,Hicks_Thermoelectric_1993,Humphrey_Reversible_2002,Humphrey_Reversible_2005,Nakpathomkun_Thermoelectric_2010,Cai_Transport_2008,Rajput_Colossal_2011,Donsa_Double_2014,Whitney_Most_2014,Whitney_Finding_2015}.

At the same time, coupled nano structures have gained increasing attention. They enable the partition of the system in a conductor and an environment whose interaction can be tailored by choosing appropriate nanostructures. 
For instance, in conductors coupled via Coulomb interaction the behavior of charge carriers becomes correlated \cite{Molenkamp_Scaling_1995,Chan_Strongly_2002,Hubel_Two_2007} which becomes evident in Coulomb drag effects \cite{Mortensen_Coulomb_2001,Shinkai_Bidirectional_2009,Levchenko_Coulomb_2008,Moldoveanu_Coulomb_2009,Sanchez_Mesoscopic_2010,Stark_Coherent_2010,Laroche_Positive_2011,Bischoff_Measurements_2015,Kaasbjerg_Correlated_2016} and the correlation of noise \cite{McClure_Tunable_2007,Goorden_Two_2007,Goorden_Cross-Correlation_2007,Michalek_Dynamical_2009,Gattobigio_Enhancement_2002,Sanchez_Electron_2008,Hussein_Heat_2016}, and are relevant for non-equilibrium fluctuation theorems~\cite{Astumian_Reciprocal_2008,Bulnes_Fluctuation_2011,Sanchez_Mesoscopic_2010}.
These days, Coulomb coupled conductors are also being used frequently as highly sensitive charge detectors~\cite{Gustavsson_Counting_2006,Fujisawa_Bidirectional_2006,Kung_Irreversibility_2012}. 
Clearly, an interesting question that follows from these developments is how coupling and correlated carrier behaviour influences the connection of heat and charge flow? Can such structure give rise to new thermoelectric effects? 

In recent years this question has gained significant attention in both theory and experimental works.  
On the one hand it leads to fundamental thermodynamic problems, that can be addressed at the mesoscopic scale such as the role of entropy production in feedback-controlled setups~\cite{Strasberg_Thermodynamics_2013,Koski_OnChip_2015} or work generation out of non thermal states~\cite{Whitney_Thermoelectricity_2016}. On the other hand, a detailed understanding of the relation between heat and charges has triggered new developments for thermal management and thermotronics for example by proposing and realizing the thermal counterparts of basic electronic devices such as thermal rectifiers \cite{Terraneo_Controlling_2002,Chang_Solid_2006,Scheibner_Quantum_2008,Ruokola_Single_2011,Matthews_Thermally_2012,Tseng_Heat_2013,Tseng_Rectification_2013,Sanchez_Heat_2015} and thermal transistors \cite{Yigen_Wiedemann_2013,Ben_Near_2014,Jiang_Phonon_2015} but also heat pumps~\cite{Arrachea_Heat_2007,Rey_Nonadiabatic_2007,Juergens_Thermoelectric_2013} and refrigerators~\cite{Edwards_Cryogenic_1995,Prance_Electronic_2009,Venturelli_Minimal_2013,Pekola_Refrigerator_2014,Feshchenko_Experimental_2014}. 

An especially interesting route points towards multi-terminal thermoelectrics. Such geometries allow a system to respond to a temperature bias with a charge flow between two other terminals which are, for example, both at a lower temperature. This means that the intimate coupling between heat and charge flow which is inherent to today's Seebeck-based devices is broken up and new thermoelectric effects are expected to arise. A large body of mostly theoretical work has investigated this field in great depth, including different mechanisms for the coupling to the heat source: electron-electron~\cite{Sanchez_Optimal_2011,Sothmann_Rectification_2012}, electron-phonon \cite{Entin_Three_2010,Jiang_Thermoelectric_2012,Jiang_Hopping_2013}, magnetic~\cite{Sothmann_Magnon_2012} or electron-photon interactions~\cite{Rutten_Reaching_2009,Ruokola_Theory_2012,Bergenfeldt_Hybrid_2014,Henriet_Electrical_2015,Hofer_Quantum_2016}, and the application of a magnetic field~\cite{Entin_Three_2012,Brandner_Strong_2013}. A plethora of configurations can be considered including quantum dots~\cite{Sanchez_Optimal_2011,Jordan_Powerful_2013,Mazza_Thermoelectric_2014,Sothmann_Thermoelectric_2015}, mesoscopic cavities~\cite{Sanchez_Thermoelectric_2011,Sothmann_Rectification_2012}, molecular junctions \cite{Entin_Three_2010}, quantum wells~\cite{Sothmann_Powerful_2013}, p-n junctions~\cite{Jiang_Three_2013}, nanowires~\cite{Bosisio_Nanowire_2016}, superlattices \cite{Choi_Three_2016}, normal-superconductor junctions~\cite{Mazza_Separation_2015} and quantum Hall edge states~\cite{Sothmann_Quantum_2014,Sanchez_Chiral_2015,Hofer_Quantum_2015,Sanchez_Effect_2016}. 
Also the number of experimental tests is increasing \cite{Matthews_Thermally_2012,Reddy_Thermoelectricity_2011}. Moreover, the acquired insight has lead to a number of promising proposals for highly efficient nanoscale energy harvesters some of which have already been demonstrated experimentally \cite{Hartmann_Voltage_2015,Roche_Harvesting_2015,Pfeffer_Logical_2015}.

One particular proposal suggests to use two Coulomb coupled quantum dots in the Coulomb blockade regime in a three terminal geometry \cite{Sanchez_Optimal_2011}. Recent experiments on such a system~\cite{Thierschmann_ThreeTerminal_2015,Thierschmann_Thermal_2015}, and the corresponding theory, are the subject of this review. The device we will consider here consists of a heat reservoir which can exchange electrons with one of the dots. The other dot connects to two reservoirs at a lower temperature, the conductor system. Energy flow is enabled through Coulomb interaction between the QDs. 
After presenting the experimental techniques and the theoretical model used in section 2, we will discuss two regimes that have been observed in these devices. The first one arises when coupling to the heat reservoir is weak so that energy transfer takes place mainly between the two QDs while heat flow from the hot reservoirs into the conductor is not dominant (section 3). In this case an the temperature bias leads to a modification of transport in the conductor system by what we call \textit{thermal gating}. This effect can be used to control charge flow in the conductor by means of temperature. Furthermore, we will present a brief discussion based on model calculations in order to substantiate that thermal gating may also be useful to design an all-thermal transistor.
The second regime will be addressed in section 4. Here, heat flow between the reservoirs is essential. We will discuss experiments that demonstrate the conversion of heat flow into a directed charge current after the proposal in Ref.\cite{Sanchez_Optimal_2011}. In this new type of heat engine electron-hole and left-right symmetry are broken by asymmetric and energy-dependent tunneling coefficients in the conductor system. In the experiments this is achieved by manipulating the potential barriers via external gate voltages. This allows for direct tests of the underlying theory and shows how the directions of heat and charge flow become decoupled in the device.

\section{Coulomb coupled quantum dots with three terminals}

\begin{figure}
	\centering
	\includegraphics[width=0.8\linewidth]{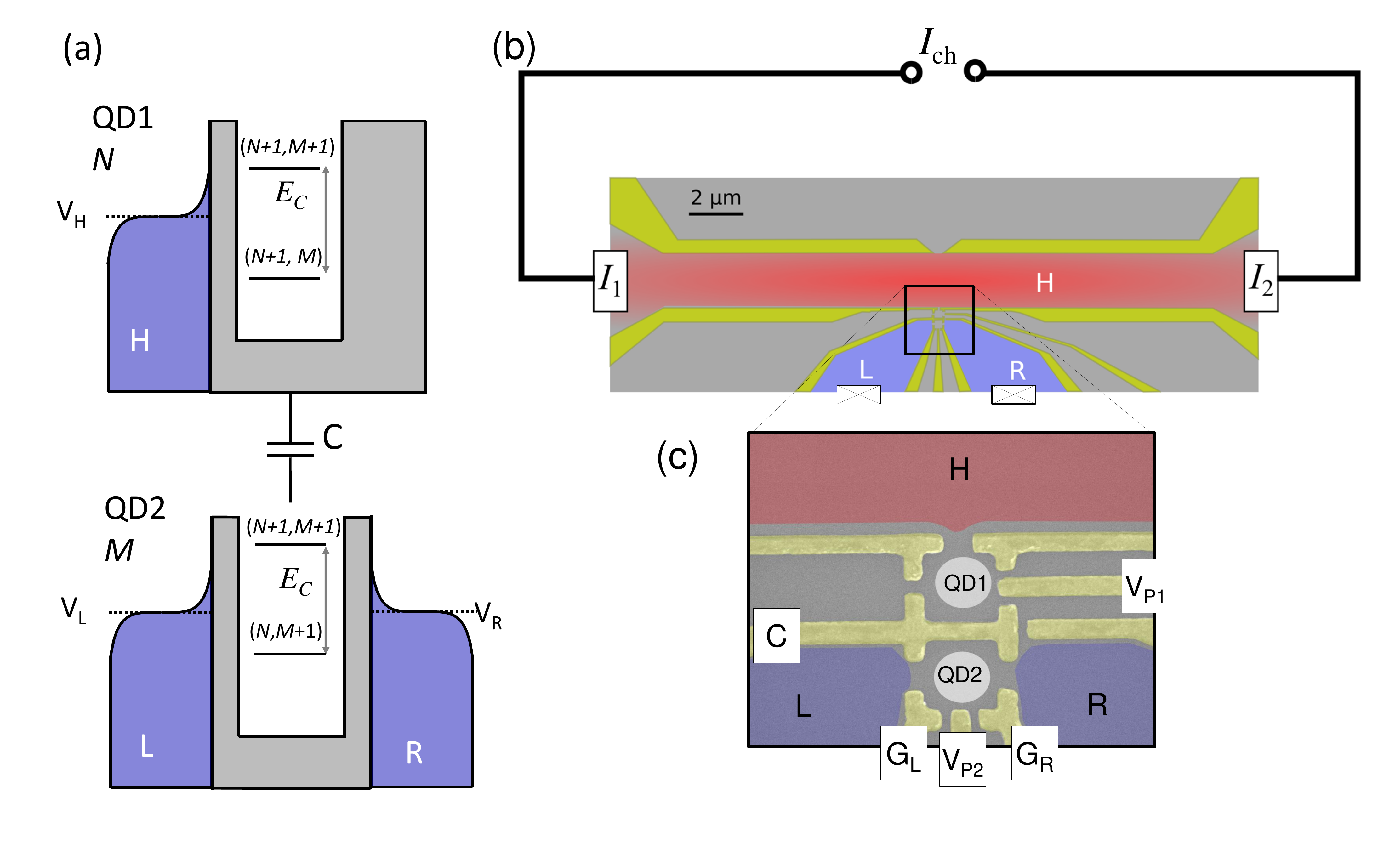}
	\caption{(a) Energy diagram of a Coulomb coupled QD system (QD1 and QD2) with three terminals H, L and R with temperatures $T_ {\rm H},$ $T_ {\rm L}$ and $T_ {\rm R}$ and electro-chemical potentials $V_\rmH$, $V_{\rm L}$ and $V_{\rm R}$, respectively. Due to capacitive coupling the electro-chemical potentials $\mu^{(1)}$ and $\mu^{(2)}$ of the dots (solid lines) change by $E_{\rm C}$ when the occupation number of the respective other dot changes by 1. (b) and (c) Experimental realization of a Coulomb coupled QD system with split gates (yellow) for thermoelectric experiments. QD1 connects to the hot reservoir H (red) while QD2 is coupled to two cold reservoirs L and R (blue). The temperature in H is controlled by driving a heating current $I_{\rm ch}$ through the channel between contacts $I_1$ and $I_2$. The Quantum point contact (QPC) is used for thermometry.}
	\label{Fig1}
\end{figure}

A three terminal geometry permits the spatial separation of the electronic conductor and the heat source. Two terminals, that we label L and R, support the electronic response to voltage or longitudinal temperature gradients: $\Delta V=V_\rmL-V_\rmR$, $\Delta T_\rmL$, $\Delta T_\rmR$. The third terminal is not invasive from the electronic point of view. Being coupled to the external heat source, at a temperature $T_\rmH$, it injects a heat current $J_\rmH$ but no electron into the conductor.  

The thermoelectric response relies on the properties of the mesoscopic region which connects the three terminals. On one hand, it is required that the symmetries of the charge conducting part can be tuned, in particular left-right and particle-hole symmetries. On the other hand, it determines what kind of interaction couples it to the heat source. For both reasons, coupled quantum dot interfaces are beneficial. Among their peculiar electronic properties, they have a discrete spectrum, with the position of the energy levels and their coupling to the lead being tunable by means of gate voltages. Also, due to their reduced dimensions, Coulomb interactions are strong. One then accesses the Coulomb blockade regime where the dynamics is governed by single-electron tunneling~\cite{single-electron}.

Let us consider an interface consisting of two dots: one is connected to the two conducting terminals, the other one tunnel-coupled to the heat source, cf. Fig.~\ref{Fig1}. The capacitive coupling between them introduces a mechanism for no particle but energy exchange between the conductor and the heat source. Fluctuations of the charge in one of the quantum dots translates into voltage fluctuations in the other one, $\delta Q=C\delta V$. Two tunneling events occurring in the same quantum dot just before and just after one of such fluctuations hence occur at different electrochemical potentials, their difference being carried by  the tunneling electron. The resulting energy transfer mediates the  injection of heat from the source which can be thus controlled at the level of single-electron processes.  

Due to the discretization of energy levels in quantum dots the amount of transferred energy in each of these processes is fixed and determined by the geometrical capacitance of the total system: $E_{\rm C}$. This quantity serves as the quantum of transferred heat. It permits the mapping of charge fluctuations into energy exchange in full counting statistics measurements~\cite{Sanchez_Detection_2012}. This way, not only the flow of heat flows but also its fluctuations can be measured~\cite{Sanchez_Correlations_2013}.

\subsection{Experimental setup}
Figure~\ref{Fig1} (b) and (c) show an experimental realization of a three-terminal device with two Coulomb coupled Quantum dots, suitable for thermoelectric experiments. It is based on split-gate technology on GaAs/AlGaAs heterostructures. Such heterostructures host a high mobility two-dimensional electron system (2DES) a few nanometers below the surface. A pattern of metallic electrodes (so-called \textit{gates}, marked yellow in Fig.~\ref{Fig1}) is brought onto the sample surface by means of e-beam and optical lithography. The 2DES can be depleted locally by applying negative voltages to the gate electrodes with respect to the electron system. In this manner insulating regions are formed according to the gate pattern. Devices realized with this technology are advantageous in that they allow \textit{in-situ} control over many system parameters. For example, by adjusting the gate voltages the thickness of each tunnel barrier and the energy of each QD can be varied individually even during the experiment.

\begin{figure}
	\centering
	\includegraphics[scale=0.7]{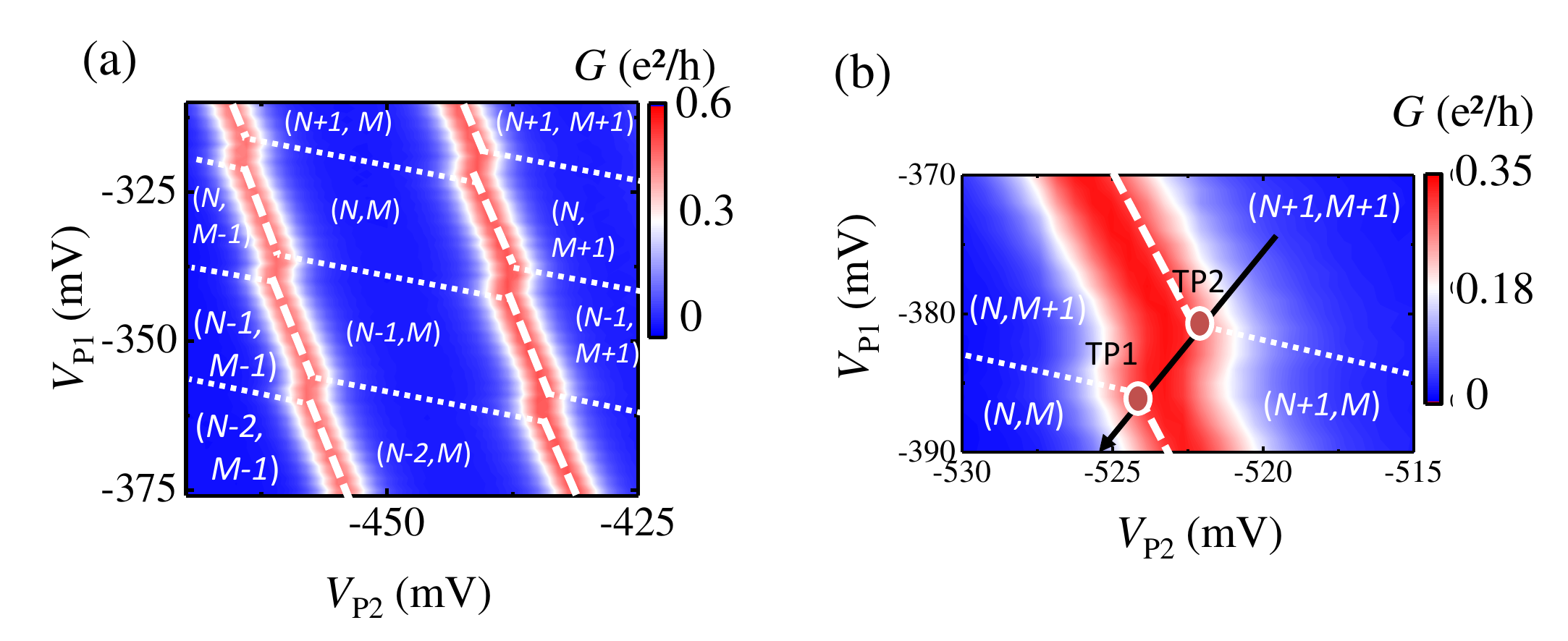}
	\caption{\textbf{(a)} Stability diagram displaying the conductance $G$ of QD2 as colour plot, obtained from experiment. The borders of the stability regions are indicated with dashed and dotted lines. \textbf{(b)} Close-up of a stability vertex. $E_{\rm C}$ can be determined from the separation of the TPs (white circles).The arrow indicates the axis of total energy.}
	\label{CQDStability}
\end{figure}

In Fig.~\ref{Fig1}(c) the QD system is shown in more detail. It is denoted QD1 and QD2, identifying each QD, respectively. QD1 can  exchange electrons only with the hot reservoir H, while QD2 can exchange electrons with the two terminals L and R, forming the conductor system.
Gate C creates a potential barrier that separates the QDs and thus strongly suppresses electron exchange while enabling a close spatial separation between the dots. The resulting inter-dot Coulomb interaction can directly be observed in the so-called stability diagram of the QD system, given in Fig.\ref{CQDStability}(a). It displays the conductance $G$ of QD2 for the $T_{\rm H}=T_{\rm L}=T_{\rm R}= 200~ \rm{mK} $ (isothermal) as a function of both QD energies, represented by the gate voltages $V_{\rm P1}$ and $V_{\rm P2}$. Resonances of $G$ due to lifted Coulomb blockade are represented with red colour. In between such resonances $G$ is suppressed. Accordingly, the occupation number $M$ of QD2 is fixed here. Capacitive inter-dot coupling becomes evident at those values for $V_{\rm P1}$ for which jumps of $G$ are observed. Here, the electron number $N$ of QD1 changes. This leads to transfer of the energy amount $E_{\rm C}$ to QD2 and thus to a sudden change of its energy. Accordingly, the regions between two conductance jumps along $V_{\rm P1}$ can be identified with fixed $N$. In this manner the diagram is divided into regions of stable charge configuration $(N,M)$. At the transition from one stability region to another the occupation number of QD1 or QD2 fluctuates. For those configurations where three stability regions meet both occupation numbers, $N$ and $M$, can change (cf. Fig.\ref{CQDStability}(b)). These points are called \textit{triple points} (TPs). The separation of two neighboring TPs in the stability diagram is determined by the inter-dot Coulomb-coupling. Hence, it can be used to measure $E_{\rm C}$ \cite{VanderWiel_Electron_2002}. For the devices discussed here $E_{\rm C}$ is typically of the order of $70-100~ \mu$V. The region around two neighbouring TPs will be called \textit{stability vertex}.

In order to conduct thermoelectric experiments at the nano-scale, precise control over the electronic temperature of the heat bath is essential while the other electron reservoirs are to remain at a lower temperature. A good way to achieve this is to apply the heating-current technique \cite{Molenkamp_Quantum_1990}. A small current $I_{\rm ch}$ that is passed through a narrow 2D channel at low temperature heats up the electron gas locally due to electron-electron scattering. In Fig.~\ref{Fig1}(a) the heating channel is denoted reservoir H (red). At both ends it opens up into large contact reservoirs, labelled $I_1$ and $I_2$, respectively. Here, heat is dissipated into the lattice due to electron-phonon interaction which cools the electron gas. Hence, by adjusting $I_{\rm ch}$ the temperature in the channel, $T_ {\rm H}$, can be controlled conveniently. $T_ {\rm H}$ can be enhanced over a wide range from typically $\Delta T < 10$ {mK up to } $\Delta T \approx 150$ mK at dilution refrigerator temperatures. $\Delta T$ can be determined quantitatively by thermopower thermometry of the quantum point contact \cite{Molenkamp_Quantum_1990} which is positioned at one side of the channel [cf. Fig.~\ref{Fig1}(a)]. When $I_{\rm ch}$ is modulated with a low frequency $\omega$ the temperature in the channel exhibits a 2$\omega$ periodic oscillation because the heating power $P$ relates to the heating current as $P\propto I_{\rm ch}^{2}$. This is another powerful feature of the current heating technique because it provides a distinct signature to effects that are driven by the channel temperature, thus allowing for convenient lock-in detection.

\subsection{Model}
The properties of such a system with Coulomb interacting QDs are well captured with the following theoretical model. Close to a stability vertex, the system can be described in terms of four charge states: $(N,M)$, $(N+1,M)$, $(N,M+1)$, $(N+1,M+1)$, depending on whether there is one or no extra charge in any of the dots. As intradot Coulomb interaction is a much larger energy scale, we can disregard configurations with two extra electrons in the same QD. 

The electrostatic potential in one of the quantum dots is sensitive to the occupation of the other one, increasing an amount $E_{\rm C}$ given by the effective capacitive couplings: $E_{\rm C}=2e^2C/(C_sC_g-C^2)$. $C_s$ and $C_g$ are the total capacitances of the system and gate dot, respectively. Every energy exchange between the two dots will be in terms of this quantity.

In the weak coupling regime, transport is due to the sequential tunneling of single electrons. The dynamics of the system is then described by a rate equation for the charge occupation probability, $P(n,m)$~\cite{Beenakker_Theory_1991,Beenakker_Theory_1992}:
\be
\dot{P}(n,m)=\sum_k\left[\Gamma^{(n{,}m){\leftarrow}(n'{,}m')}_kP(n',m')-\Gamma^{(n'{,}m'){\leftarrow}(n{,}m)}_kP(n,m)\right].
\ee
The tunneling rates depend on the Fermi distribution function of lead $k$, $f_k(E)=[1+\exp(E/\kT_k)]^{-1}$ and on the transparency $\Gamma_k^{(i)}$. They are in general energy-dependent, and hence are different for processes occurring through the same barrier $k$ at different occupations of the respective other dot, indicated by $i=0,1$. For electrons tunneling in the dots:
$\Gamma_k^{(1,i){\leftarrow}(0,i)}=\Gamma_k^{(i)}f_k(\Delta U_2^{(i)}-eV_k)$ for QD2 with $k = L,R$ and  $\Gamma_H^{(i,1){\leftarrow}(i,0)}=\Gamma_H^{(i)}f_H(\Delta U_1^{(i)}-eV_H)$ for QD1,
with $\Delta U_\alpha^{(i)}$ being the change in the electrochemical potential during the tunneling process in dot $\alpha$. It depends on the position of the level and voltages~\cite{Sanchez_Optimal_2011}. Note that $\Delta U_\alpha^{(1)}=\Delta U_\alpha^{(0)}+E_{\rm C}$. For tunneling-out electrons ($\Gamma_k^{(0,i){\leftarrow}(1,i)}$ and $\Gamma_H^{(i,0){\leftarrow}(i,1)}$), one needs to replace $f_k\rightarrow 1-f_k$ in the above expressions.

The charge and heat currents are obtained with the stationary solution of the  rate equations, $\dot{P}(n,m)=0$: 
\be
I_k=e\sum_m\left[\Gamma_k^{(0,m){\leftarrow}(1,m)}\bar{P}(1,m)-\Gamma_k^{(1,m){\leftarrow}(0,m)}\bar{P}(0,m)\right],
\ee
for charge and
\begin{align}
J_k&=\sum_m(\Delta U_2^{(m)}-eV_k)\left[\Gamma_k^{(0,m){\leftarrow}(1,m)}\bar{P}(1,m)-\Gamma_k^{(1,m){\leftarrow}(0,m)}\bar{P}(0,m)\right]\\
J_\rmH&=\sum_n(\Delta U_1^{(n)}-eV_H)\left[\Gamma_k^{(n,0){\leftarrow}(n,1)}\bar{P}(n,1)-\Gamma_k^{(n,1){\leftarrow}(n,0)}\bar{P}(n,0)\right],
\end{align}
for heat. Obviously, $I=I_L=-I_{\rm R}$ (by  charge conservation) and $I_H=0$. Particle and heat currents are defined positive when flowing into the leads.
\begin{figure}
	\centering
	\includegraphics[width=0.7\linewidth]{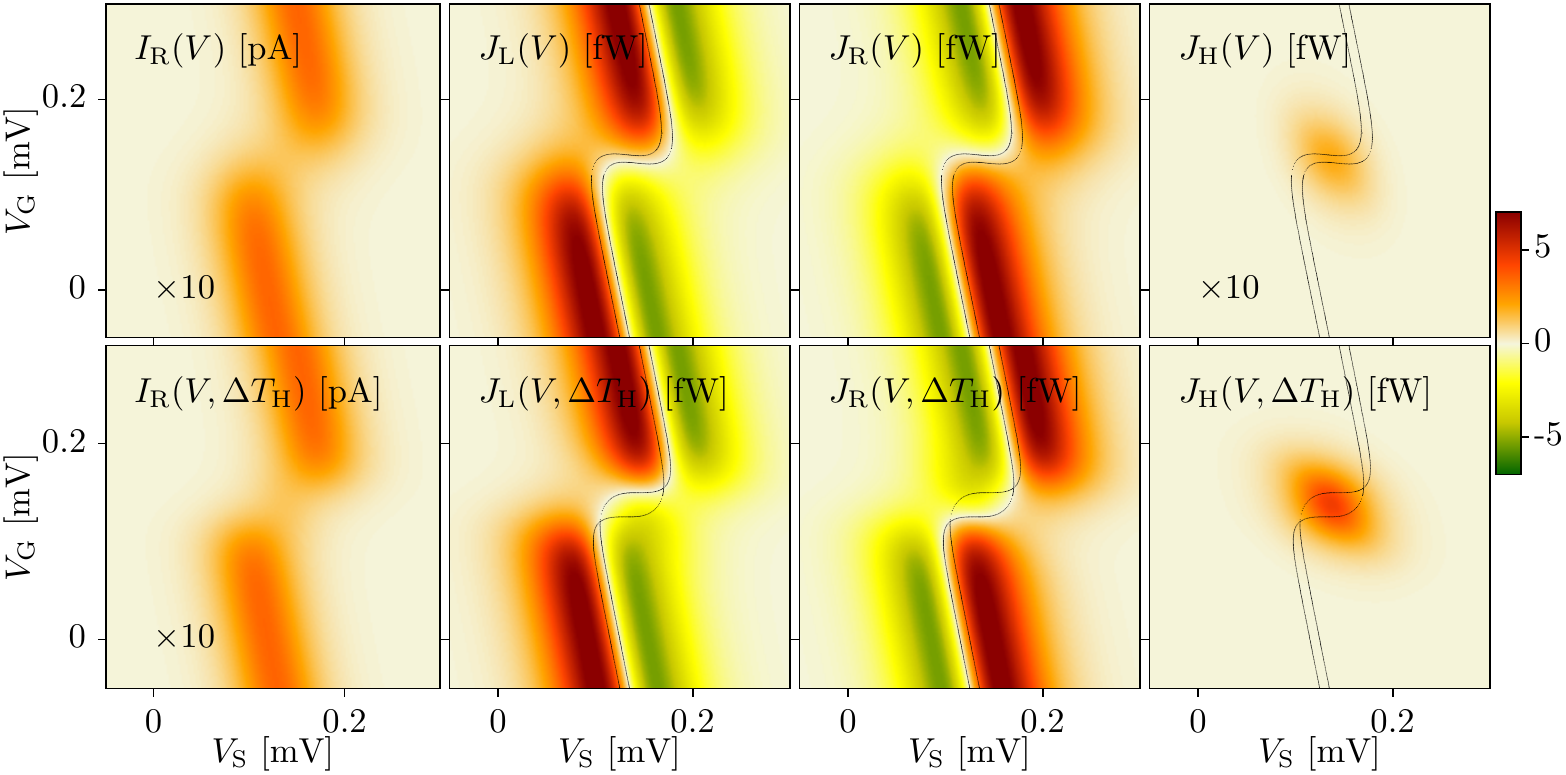}
	\caption{Top: Charge and heat currents as functions of the applied gate voltages to the system and gate dots, for an applied bias voltage $\Delta V_{\rm LR} = 10~\mu$V at $T=0.243~$K. We consider symmetric couplings: $\Gamma_{L}^{(i)}=\Gamma_{R}^{(i)}=23.2~\mu $eV, $\Gamma_{H}^{(1)}=10~\mu $eV, $E_{\rm C}=90~\mu$eV. The charge current peaks around the degeneracy of configurations $(N,M)$ and $(N{+}1,M)$. There, the heat currents change sign (marked with dotted lines), as expected for single level quantum dots. Note that they cannot be zero simultaneously due to Joule heating. At the center of the stability vertex, charge fluctuations are strong in both dot, resulting in a finite heat current $J_\rmH$. Bottom: Same as top, but having a cold gate dot with $\Delta T_\rmH=-0.06~$K. The gate reservoir serves as a heat drain.}
	\label{currents}
\end{figure}

In Fig.~\ref{currents}, we plot the charge and heat currents in the presence of a finite bias voltage at the isothermal case $\Delta T_k=0$. Along the degeneracy of configurations with $N$ and $N{+}1$ electrons in the conductor dot, the charge current shows a typical Coulomb blockade peak, in agreement with the experimental results in Fig.~\ref{CQDStability}(c). The Peltier-like heat current $J_\rmL$ and $J_\rmR$ presents a saw-tooth pattern, a well known result for single-level quantum dots~\cite{Beenakker_Theory_1992,Staring_Coulomb_1993}. Each of them vanishes when the quantum dot level aligns with the corresponding lead chemical potential. Note that there is no condition for which $J_\rmL=J_\rmR=0$ due to Joule heating.

These features are shifted around the stability vertex, where fluctuations of the charge in both dots coexist. This gives rise to enhanced correlations which are responsible for the effective coupling of the two subsystems~\cite{Sanchez_Correlations_2013} and ultimately of the performance of the device as a heat engine~\cite{Sanchez_Optimal_2011,Thierschmann_ThreeTerminal_2015}. Only in this region, a finite heat current is absorbed by QD1 (the gate dot), as shown in Fig.~\ref{currents}. Away from it, cross-correlations of the charge fluctuations are supressed and hence the two systems are effectively uncoupled. This effect has been interpreted as an autonomous feed-back control mechanism for which the gate dot can work as a Maxwell demon~\cite{Strasberg_Thermodynamics_2013,Koski_OnChip_2015}

Indeed, as shown in Fig.~\ref{currents}, if the gate dot is cold, the conductor heat currents are almost unaffected far from the stability vertex. However as the fluctuations in the gate dot start to be important, the gate dot starts acting as a heat sink. Then, the lines where the heat currents are zero in the conductor cross at two points. There, all the Joule heating flows into the gate reservoir, i.e. current flows without heating up the conductor. In the region between them, both  leads in the conductor are cooled simultaneously. The gate dots behaves then as a Maxwell demon~\cite{Koski_OnChip_2015}.

The lock-in technique used in the experiments measures the current at an increased gate temperature $T_\rmH=T+\Delta T_\rmH$ on top of a background given by having a cold reservoir, $T_\rmH=T$. Hence, the experimentally relevant quantity is:
\be
\Delta I=I(V,\Delta T_\rmH)-I(V,0).
\label{eq:thgat}
\ee

\section{Thermal gating}

The following section will discuss how the transport properties of Coulomb coupled QDs are affected when a temperature difference is applied. We will find a significant response even without considerable heat flow between the hot and the cold reservoirs. This is related to the fact that on the one hand the rate of occupation fluctuations of a QD is related to the temperature of the adjacent reservoirs. On the other hand, mutual Coulomb interaction of two QDs couples the individual QD energies to the occupation number of the respective other QD. Hence, for the system shown in Fig.~\ref{Fig1} the temperature in reservoir H influences (via occupation fluctuations on QD1) the energy levels of QD2 which, in turn, changes transport between reservoirs L and R. Recent experiments \cite{Thierschmann_Thermal_2015} have shown that this long-range interaction indeed leads to a sizeable effect called \textit{thermal gating}. It can be used to control, for example, charge flow across QD2 by means of temperature in reservoir H. 

\begin{figure}
	\centering
	\includegraphics[width=0.9\linewidth]{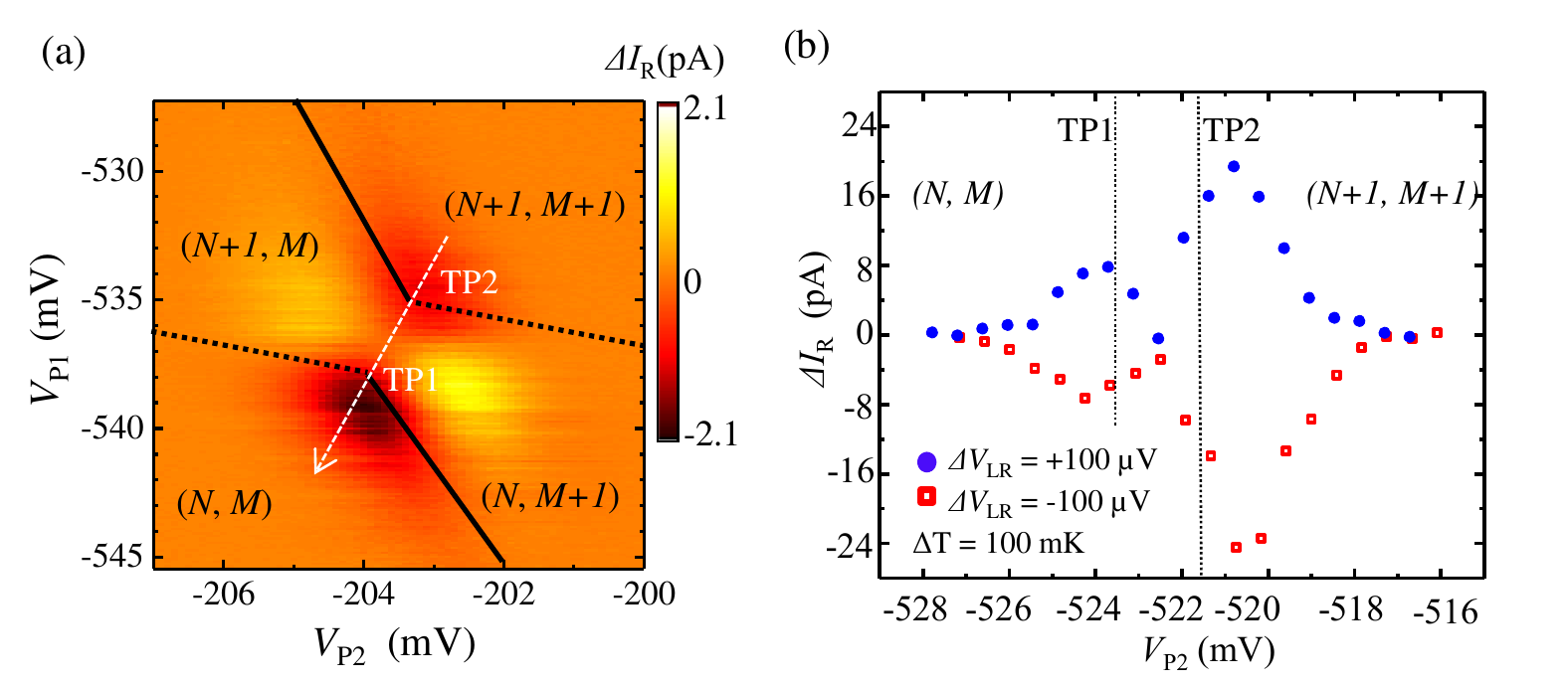}
	\caption{Thermal gating of charge currents. (a) $\Delta I_{\rm R}$ for an increase of $T_ {\rm H}$ by $\Delta T \approx 100$ mK at a stability vertex with $\Delta V_{\rm LR} \approx 10~\mu$V. The borders of the stability regions are indicated with black, solid and dotted lines. The white arrow indicates the axis of total energy. (b) $\Delta I_{R}$ obtained along the axis of total energy for $\Delta T \approx 100 $ mK from a feature similar to that in (a). The data are plotted against V$_{P2}$. The positions of TP1 and TP2 are indicated. Blue circles: $\Delta V_{\rm LR} = 100~\mu$V. Red squares: $\Delta V_{\rm LR} = -100~\mu$V.}
	\label{ThermalGating_1}
\end{figure}

In order to demonstrate thermal gating a finite voltage $\Delta V_{\rm LR}$ of several 10 $\mu$V is applied between reservoirs L and R in the experiments reported in \cite{Thierschmann_Thermal_2015} on a device similar to the one shown in Fig.\ref{Fig1}. $T_\rmH$ is increased by $\Delta T \approx 100$ mK by means of ac-current heating. Measuring the $2\omega$-current in reservoir R with a lock-in provides direct access to the change of current, $\Delta I_{\rm R}$, due to the change of $T_\rmH$.

Results obtained in this manner ($\Delta V_{\rm LR}=-10~\mu{\rm eV}$) are depicted in Fig.\ref{ThermalGating_1}(a).
A change of $I_{\rm R}$ is observed around the stability vertex (indicated by dashed and solid lines). Interestingly, the impact of a temperature increase is not uniform: In the regions $(N{+}1,M) $and $(N, M{+}1)$ an enhancement of current is observed ($\Delta I_{\rm R} > 0$) while for $(N,M)$ and $(N{+}1,M{+}1)$, $I_{\rmR}$ becomes reduced ($\Delta I_{\rm R} <0$). This gives rise to a characteristic clover-leaf shaped pattern.  

Fig.~\ref{ThermalGating_1}(b) compares $\Delta I_\rmR$ of a similar feature for opposite bias voltage $\Delta V_{\rm LR} = \pm 100~\mu$eV along the trace cutting through both TPs, as indicated by a dotted line in Fig.~\ref{ThermalGating_1}(a). Along this line both QDs are tuned in energy simultaneously, which is why this axis is sometimes called \textit{axis of total energy} \cite{VanderWiel_Electron_2002}.
Note that in Fig.~\ref{ThermalGating_1}(b) $\Delta I_{\rm R}$ is plotted as a function of gate voltage $V_{\rm P2}$. 
For negative (squares) and positive (circles) bias voltages a maximum signal as large as $\Delta I_{\rm R} = \pm  20$ pA is observed. The traces peak closely beyond the TPs. At the center between the TPs and towards the centers of the charge stability regions the signal becomes suppressed.

\begin{figure}[t]
	\centering
	\includegraphics[width=0.8\linewidth]{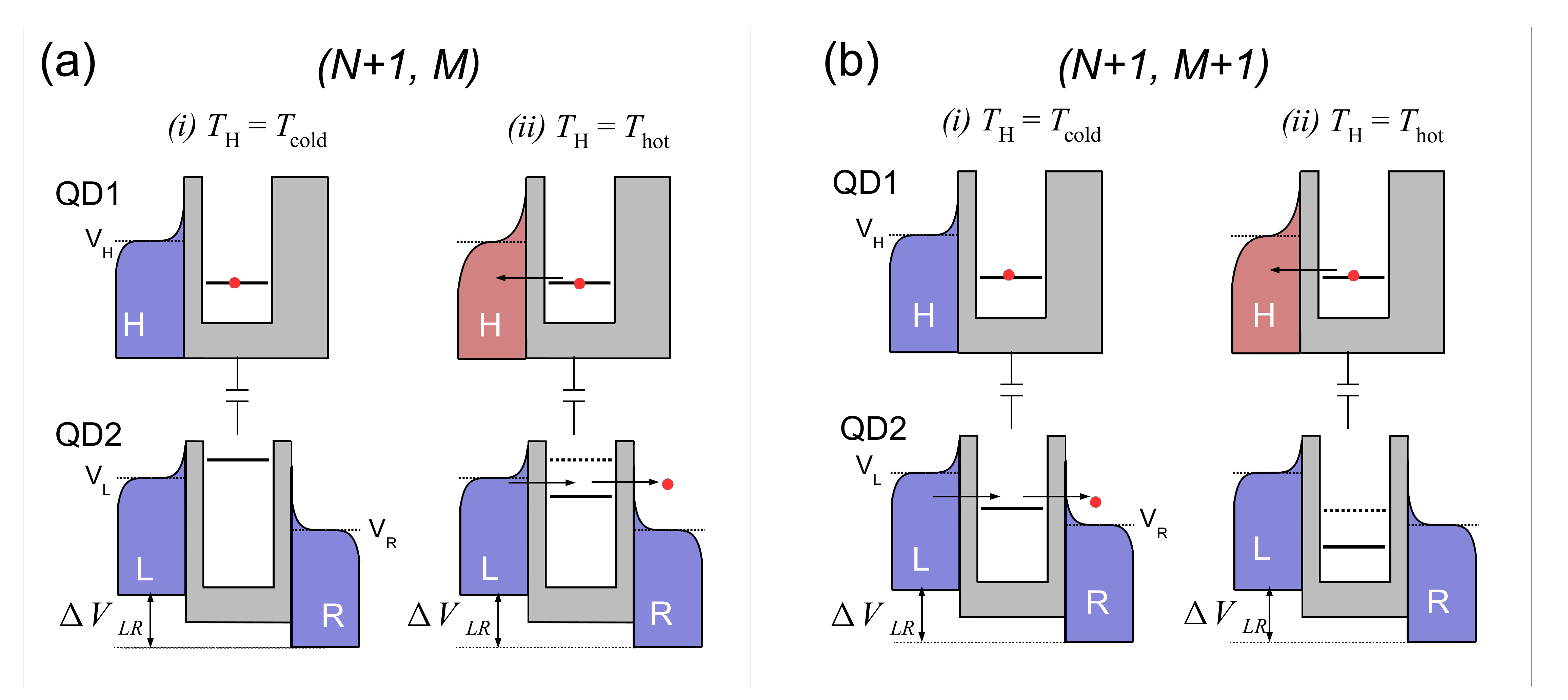}
	\caption{Schematic energy diagram of the QD-system for the stability regions (a) $(N+1,M)$ and (b) $(N+1,M+1)$ with $\Delta V_{\rm LR}<0$. Each configuration is shown for \textit{(i)} low and \textit{(ii)} high $T_{\rmH}$. For QD2 the solid line indicates the enabled, the dotted line the suppressed transport channel. Arrows indicate increased charge fluctuation.}
	\label{fig:ThermTran:cartoon}
\end{figure}

In order to better understand thermal gating on a microscopic level, we shall consider the alignment of the individual QD energy levels in the regions surrounding a stability vertex in more detail.
Figure~\ref{fig:ThermTran:cartoon}~(a) shows the QD system for the stability region $(N+1,M)$ with $\Delta V_{\rm LR} < 0$. We emphasize that due to the ac-current heating $T_\rmH$ oscillates in the experiments. Therefore, the measured signal $\Delta I_{\rm R}$ is the difference of (\textit{i}) $I_{\rm R}$ at $T_\rmH= T_{\rm cold}$ and (\textit{ii}) $I_{\rm R}$ at $T_\rmH= T_{\rm cold} + \Delta T = T_{\rm hot}$. The first case (\textit{i}) is sketched on the left side of Fig.~\ref{fig:ThermTran:cartoon}~(a): the electro-chemical potential of QD1 to exhibit $(N+1)$ electrons is indicated by a solid line. It is situated below $V_{\rmH}$, i.e., the QD occupation number is fixed at $(N+1)$. Fluctuations are small because $T_{\rmH}$ is at a minimum. QD2 is occupied with $M$ electrons because charge carriers in reservoir L do not have enough energy to tunnel into reservoir R via the QD2 state $\mu^{(2)}$($N+1$,$M$+1) (solid line). Thus, no charge current flows through QD2. 
If $T_{\rmH}$ is increased to $T_{\rm hot}$ [case (\textit{ii}), right hand side in Fig.~\ref{fig:ThermTran:cartoon}~(a)] empty states are created below $V_\rmH$. This enhances the probability for the $(N+1)^{th}$ electron to hop off QD1. Therefore, the rate at which QD1 fluctuates between the states $(N+1) \leftrightarrow N $ increases. Whenever QD1 switches to the $N$-state, $\mu^{(2)}$ drops by $E_{\rm C}$. It then lies within the bias window $\Delta V_{\rm LR}$. As a result a current flows across QD2 [indicated by red arrows].
This mechanism leads to a temperature driven change of the conductance of QD2: When $T_{\rmH}$ is at a minimum, charge flow between L and R is suppressed. It becomes enabled if $T_{H}$ increases. The resulting current change $\Delta I_{\rm R}$ is then detected as positive in the experiment. 
\begin{figure}
	\centering
	\includegraphics[width=0.6\linewidth]{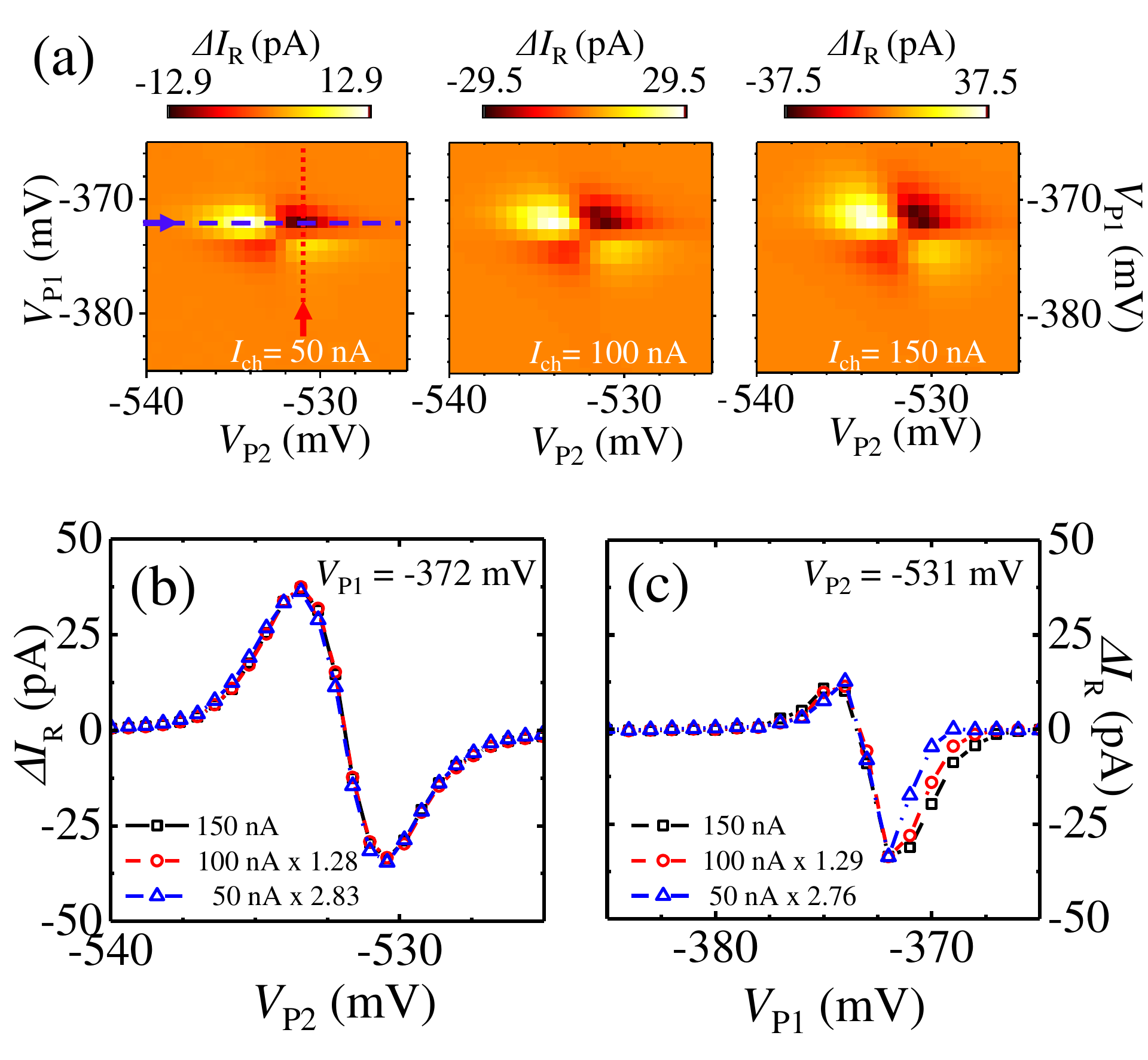}
	\caption{$\Delta I_{R}$ for different $I_{\rm ch}$. (a) Thermal gating clover leaf structure for $I_{\rm ch}=$ 50, 100 and 150 nA. The dashed, blue line indicates $V_{\rm P2}=372$ mV. The dotted, red line denotes $V_{\rm P1}=531$ mV. For these configurations data for different $I_{\rm ch}$ are compared in (b) (fixed $V_{\rm P1}$) and (c) (fixed $V_{\rm P2})$. The traces for 100 nA and 50 nA are multiplied by constant factors.}
	\label{Tdep}
\end{figure}

Let us now see what happens in the ($N+1,M+1$) region. The corresponding level alignments is depicted in Fig.~\ref{fig:ThermTran:cartoon}~(b). Starting again with the condition $T_{\rmH} = T_{\rm cold}$, i.e. with fixed $(N+1)$, we find transport across QD2 enabled because $\mu^{(2)}$($N+1$,$M$+1) lies below $V_{\rmL}$ and above $V_{\rmR}$. Hence, a current is flowing. When $T_{\rmH}$ is enhanced [right side in Fig.~\ref{fig:ThermTran:cartoon}~(b)], occupation fluctuations on QD1 tend to reduce the energy of QD2 so that the (M+1) electron becomes trapped on QD2 whenever QD1 switches to the $N$ state. In this configuration a temperature increase tends to block transport. The resulting $\Delta I_{\rm R}$ is negative. 
Similar considerations apply to the other stability regions of the vertex, $(N,M)$ and $(N, M+1)$, with the main difference that QD1 is then occupied most of the time with $N$ electrons. Fluctuations occur at energies above $V_\rmH$ when an electron is added to the dot. This introduces an additional sign change to the $\Delta T$ - $\Delta I_{\rm R}$ relation which then leads to the observed clover-leaf pattern.
Notably, a similar pattern has been observed for the cross correlation of shot noise in Coulomb coupled QD systems \cite{McClure_Tunable_2007}. This is in line with the picture for thermal gating, because the enhancement and suppression of current with $T_\rmH$ is closely related to whether occupation fluctuations on the dots correlate positively or negatively.

These experiments show that thermal gating gives rise to a considerable response of $I_\rmR$ to a change of $T_\rmH$. It is an interesting result because the two systems of different temperature are electrically disconnected and do not couple directly but only via QD1. Yet, reservoir H acts as a gate which manipulates the current between the cold reservoirs with the important feature that it is operated only thermally.

The on/off-ratio of the device reported on in Ref~\cite{Thierschmann_Thermal_2015} is of the order of 5\%. This estimate has been obtained by relating $\Delta I_{\rm R}$ for a given $\Delta T$ to the total current $I_{\rmR}$ which flows at a fixed $T_{\rmH}$. 
However, it could easily be increased further by optimizing the device layout, for example towards stronger inter-dot coupling $E_{\rm C}$. This will prevent currents from leaking through QD2 when, for example, $T_{\rmH}=T_{\rm cold}$ (off-position) and therefore will lead to more well-defined on/off states. 

The suppression or enhancement of $I_{\rm R}$ can also be augmented by increasing $\Delta T$. This is shown in Fig.~\ref{Tdep} where thermal gating is observed for different heating currents ($I_{\rm ch} = 50, 100, 150$ nA). For the series of measurements shown in Fig.~\ref{Tdep}(a) $\Delta I_\rmR$ increases by approximately a factor 2.8. Interestingly, when plotted against the energy of QD2, represented by $V_{\rm P2}$, the line shape does not change for different $I_{\rm ch}$. This becomes visible even more clearly in Fig.~\ref{Tdep} (b) where data have been extracted from Fig.~\ref{Tdep} (a). The curves can be brought into alignment by simple scaling. However, this is not possible along $V_{\rm P1}$. Here a clear change in line shape is observed for different $I_{\rm ch}$. This mirrors the fact that the line shape is determined by the Fermi-distribution in the adjacent reservoirs: Along $V_{\rm P1}$ the relevant temperature is $T_\rmH$ for which the Fermi-distribution changes with $I_{\rm ch}$. Along the $V_{\rm P2}$-axis, in contrast, the temperature in reservoirs L and R matters, which stays at $T_{\rm cold}$. This indicates that besides for manipulation of charge currents, thermal gating could also be used for non-invasive thermometry, e.g. to monitor the temperature in quantum circuits. 

\begin{figure}
	\centering
	\includegraphics[width=0.8\linewidth]{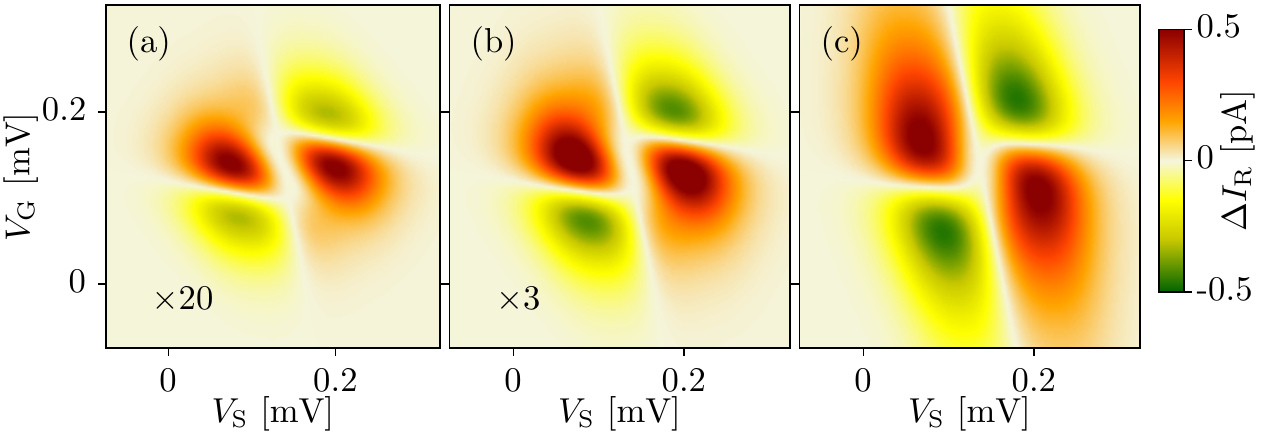}
	\caption{Theoretical $\Delta I_{R}$ as a function of the position of the quantum dot levels. Each panel corresponds to a different temperature in the gate dot: (a) $\Delta T_\rmH=0.012~$K, (b) $\Delta T_\rmH=0.122~$K, (c) $\Delta T_\rmH=0.608~$K, with $T_{\rm cold}=0.243~$K and $\Delta V_{\rm LR} = 100~\mu$V. Same parameters as in Fig.~\ref{currents}, except for $\Gamma_\rmH^{(n)}=1~\mu$eV. The size and magnitude of the clover-leaf structure increase with the gate temperature.}
	\label{TdepTheory}
\end{figure}

Simple model calculations using Eq.~\eqref{eq:thgat} reproduce this effect, as shown in Fig.~\ref{TdepTheory}. They also show that the center of the clover-leaf structure changes its shape when $k_{\rm B}\Delta T_\rmH$ gets much smaller than $eV$ and $E_{\rm C}$.

\subsection{All-thermal transistor}

The thermal gating effect suggests the possibility to modulate not only the charge but also the heat currents. In this case, it is particularly interesting to consider flows generated by a thermal gradient in the conductor. The thermal gating of transport in the system could then be used to define an all-thermal transistor \cite{Thierschmann_Thermal_2015}. This effect has been recently discussed for conductors coupled to phonons~\cite{Jiang_Phonon_2015}. 

\begin{figure}
	\centering
	\includegraphics[width=0.8\linewidth]{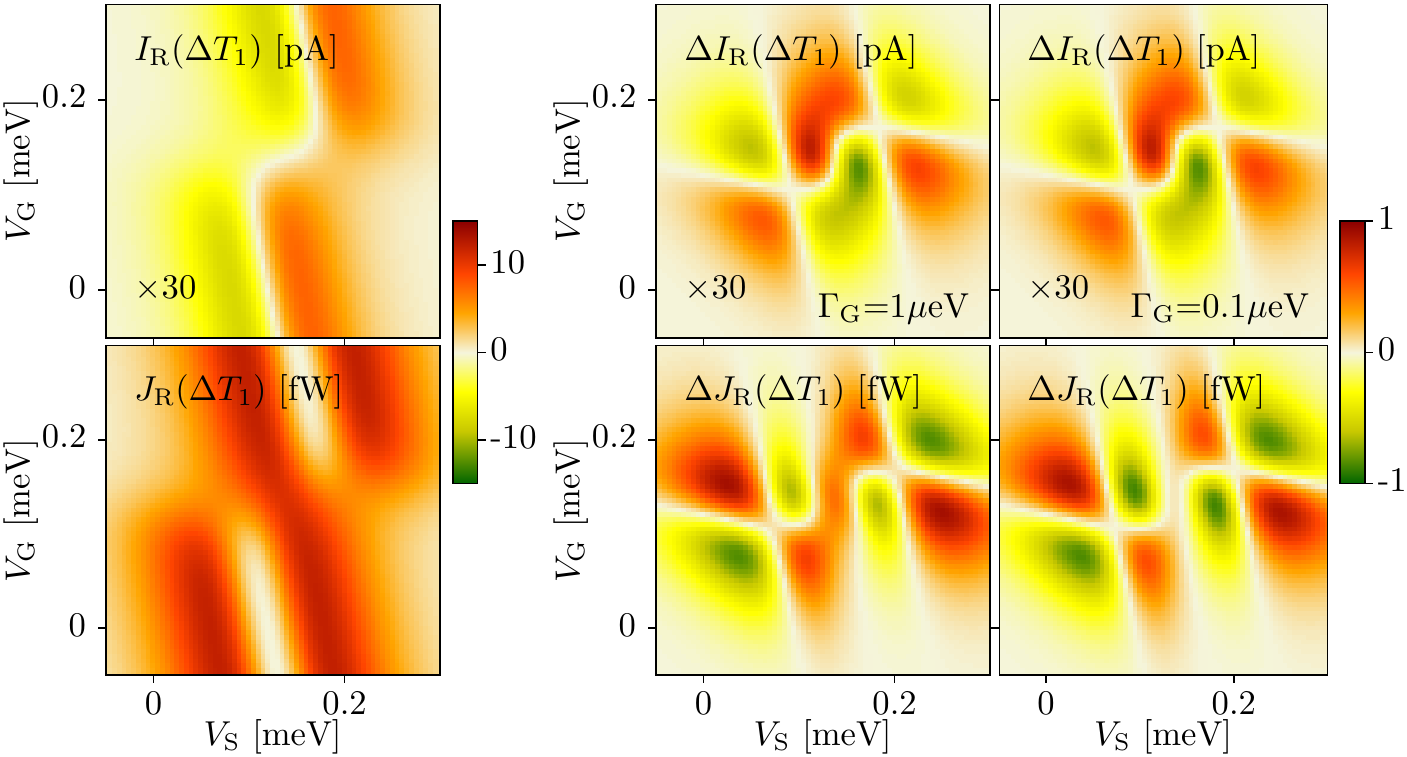}
	\caption{Thermal gating of thermally generated currents. The leftmost panels show the charge and heat currents in the right lead when introducing a temperature gradient $\Delta T_{\rm LR}=0.122~$mK, as a function of the quantum dot potentials, for $\Gamma_\rmH^{(n)}=1~\mu$eV. The charge current shows the expected saw-tooth pattern typical of the Seebeck effect for quantum dots. The heat current has a double peak which vanishes as the conducting level crosses the Fermi energy. The rightmost set of panels show the thermal gating of these currents when the temperature of the third terminal is increased. Two different tunneling rates $\Gamma_\rmH^{(n)}$ are considered, evidencing that while the effect on the charge current is unaffected, it considerably modifies the heat properties.}
	\label{ThermTansistor}
\end{figure}

In order to investigate this effect with the model introduced in section 2, we consider an enhanced temperature in the left contact compared to the right one, $\Delta T_{\rm LR}>0$. The resulting charge current changes sign at the crossing of the conducting level by the Fermi energy, as expected for the Seebeck effect in quantum dots~\cite{Beenakker_Theory_1992,Staring_Coulomb_1993,Svensson_Lineshape_2012}, even in the nonlinear regime~\cite{Svilans_Nonlinear_2016}. The heat current vanishes at the same point, but it is always positive (as required by Clausius' statement), cf. Fig.~\ref{ThermTansistor}. 

The more complicated structure of the currents (as compared to the electrically driven behaviour shown eg. in Fig.~\ref{currents}) translates into the thermal gating signals, 
\begin{align}
\Delta I_{\rm R}&=I_{\rm R}(\Delta T_\rmL,\Delta T_\rmH)-I_{\rm R}(\Delta T_\rmL,0)\\
\Delta J_k&=J_k(\Delta T_\rmL,\Delta T_\rmH)-J_k(\Delta T_\rmL,0)
\label{eq:allthgat}
\end{align}
which change sign around the maxima of the corresponding current, see Fig.~\ref{ThermTansistor}. In particular, the gating of the heat currents show a double clover-leaf structure, related to the two triple points of the stability diagram.

In order to have a proper thermal transistor leakage currents from the third terminal H should be suppressed. Obviously, heat injected from this terminal must affect the thermal currents in the conductor. The later will be larger the larger the coupling $\Gamma_\rmH$ is. In order to test this effect, we compare the thermal gating behaviour for different couplings to the third terminal in Fig.~\ref{ThermTansistor}. The charge current is essentially unaffected. For the heat current, we can distinguish two different behaviours: a feature appears in the center of the stability vertex (where $J_\rmH$ is larger) which is suppressed with smaller $\Gamma_\rmH$. This is a clear signature of heat leaking from the gate. However, the clover leaf structures around the triple points do not change. We can therefore conclude that thermal gating there does not depend on leakage currents, but rather on the non-equilibrium distribution of QD2.

\section{Energy harvesting}

In this section we will discuss in more detail the role of heat flow between reservoir H and the conductor system. As indicated in section 2, heat flow becomes relevant at the center of the stability vertex where hot and cold system are most effectively coupled. Here, a heat current is injected into the conductor system when the temperature in H is elevated.

\begin{figure}
	\centering
	\includegraphics[width=0.6\linewidth]{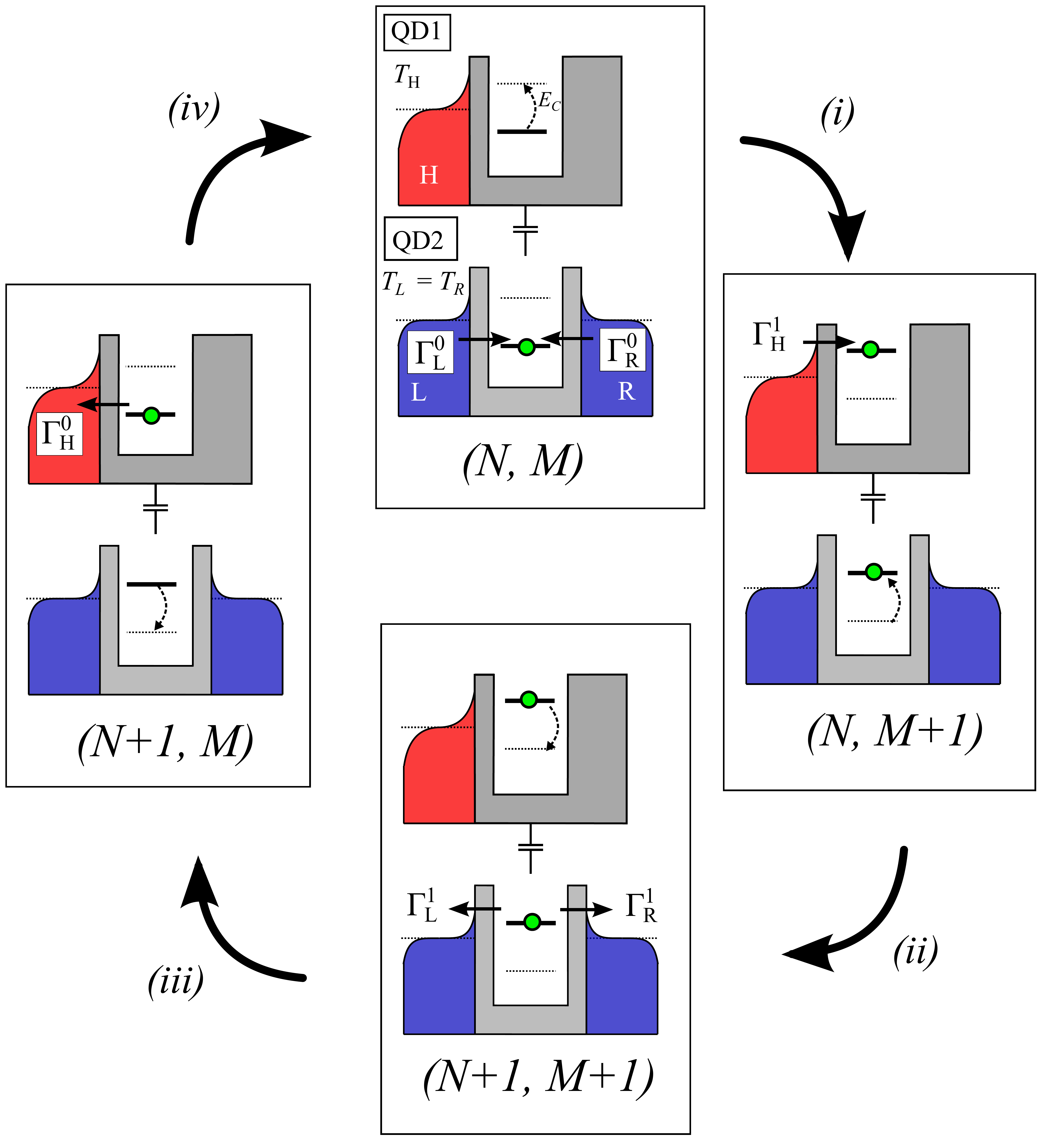}
	\caption{Heat flow cycle after Ref.~\cite{Sanchez_Optimal_2011} for $T_{\rm H}>T_{\rm L}=T_{\rm R}$. When charge fluctuations are correlated in such a way that the system undergoes the sequence $(i) \to (ii) \to (iii) \to (iv) \to (i)$ (clockwise direction), an energy package $E_{\rm C}$ is extracted from reservoir H and is delivered to the cold system. The opposite direction (counter clockwise) is suppressed because it reduces entropy. The tunneling coefficients are denoted $\Gamma^i_{j}$ where $j(=\rm L,R,H)$ denotes electron reservoir which is involved in the tunnelling process and $i(=0,1)$ indicates whether the respective other dot is occupied with an additional electron or not.}
	\label{fig:HCC:Converter_Concept}
\end{figure}

Interestingly, the heat current can be converted into a directed charge current by choosing appropriate device parameters for the conductor system, as it has been proposed recently by one of us and M. B\"uttiker \cite{Sanchez_Optimal_2011}. In order to understand the working principle of this heat-to-current conversion, let us take a closer look at how heat flow takes place on a microscopic level. Obviously, a \textit{series} of tunnelling events is needed, because a single tunnelling process exchanges energy only between the dots, but not between the reservoirs. 
In fact, a 4-step-cycle is required \cite{Sanchez_Mesoscopic_2010}. During this cycle the QD system undergoes the sequence of states $(N,M) \rightarrow (N,M+1) \rightarrow (N+1,M+1) \rightarrow (N+1,M) \rightarrow (N,M)$, as shown in Fig.\ref{fig:HCC:Converter_Concept}.
The cycle begins with \textit{(i)} an electron tunnelling from one of the cold reservoirs in the conductor system onto the QD system in the state ($N,M$) at a low energy [cf. Fig.~\ref{fig:HCC:Converter_Concept}]. This induces the transition to $(N, M+1)$. Next \textit{(ii)}, a hot electron from reservoir H tunnels onto QD1 [$(N,M+1)\rightarrow (N+1, M+1)$]. This process extracts thermal energy from the hot reservoir H which is then 'stored' in the QD system and the energy of QD2 is elevated by the amount $E_{\rm C}$. In a third step \textit{(iii)}, the electron occupying QD2 tunnels back into either of the adjacent reservoirs [$(N+1,M+1) \rightarrow (N+1, M)$]. This process delivers the stored energy package into the cold conductor system. In a final step \textit{(iv)} the electron on QD1 tunnels back into reservoir H, but now at a lower energy [$(N+1,M) \rightarrow (N, M)$]. This restores the initial state of the QD system. As a net result, the energy package $E_{\rm C}$ has been extracted from H and has been delivered to the conductor system.

The key to convert the heat flow into a charge current lies in rectifying the charge fluctuations of QD2 during step 1 and 3 of the cycle. Section 2 indicates how this can be achieved: The tunnelling rates can be controlled not only through temperature but also through the tunnelling coefficients of the corresponding potential barrier. In general, one has to consider individual coefficients $\Gamma^{(i)}_k$ for each step of the heat transfer cycle. Here $\Gamma^{(i)}_k$ indicates the coefficient for an electron tunnelling between QD2 and reservoir $k=\rmL,\rmR$ while QD1 exhibits one additional electron ($i=1$) or not ($i=0$). A particular interesting case is arises if the coefficients are asymmetric such that both left-right and electron-hole symmetry are broken. In this case one obtains $\Gamma^{(1)}_\rmL \Gamma^{(0)}_\rmR \neq \Gamma^{(1)}_\rmR \Gamma^{(0)}_\rmL$. This means that the net probability for an electron to enter QD2 from L in step 1 and to leave into R in step 3 is different from that of the opposite process, i.e. for an electron to first enter from R and to leave into L. Hence, thermally driven occupation fluctuations become rectified and a directed charge current $I$ arises which is given by \cite{Sanchez_Optimal_2011}:
\begin{equation}
I = \frac{e}{E_{\rm C}} \Lambda J_\rmH.
\label{HCC current}
\end{equation}     
Here $J_\rmH$ is the heat current into QD1, $E_{\rm C}$ is the energy package exchanged between the QDs due to Coulomb coupling and $e$ denotes the electronic charge. The parameter $\Lambda$, given by
\begin{equation}
\Lambda = \frac{\Gamma^{(1)}_\rmL \Gamma^{(0)}_\rmR - \Gamma^{(1)}_\rmR \Gamma^{(0)}_L}{(\Gamma^{(0)}_\rmL + \Gamma^{(0)}_\rmR)  (\Gamma^{(1)}_\rmL \Gamma^{(1)}_\rmR)},
\label{Lambda}
\end{equation}
describes the tunnelling asymmetry in the conductor system.

\begin{figure}
	\centering
	\includegraphics[width=0.9\linewidth]{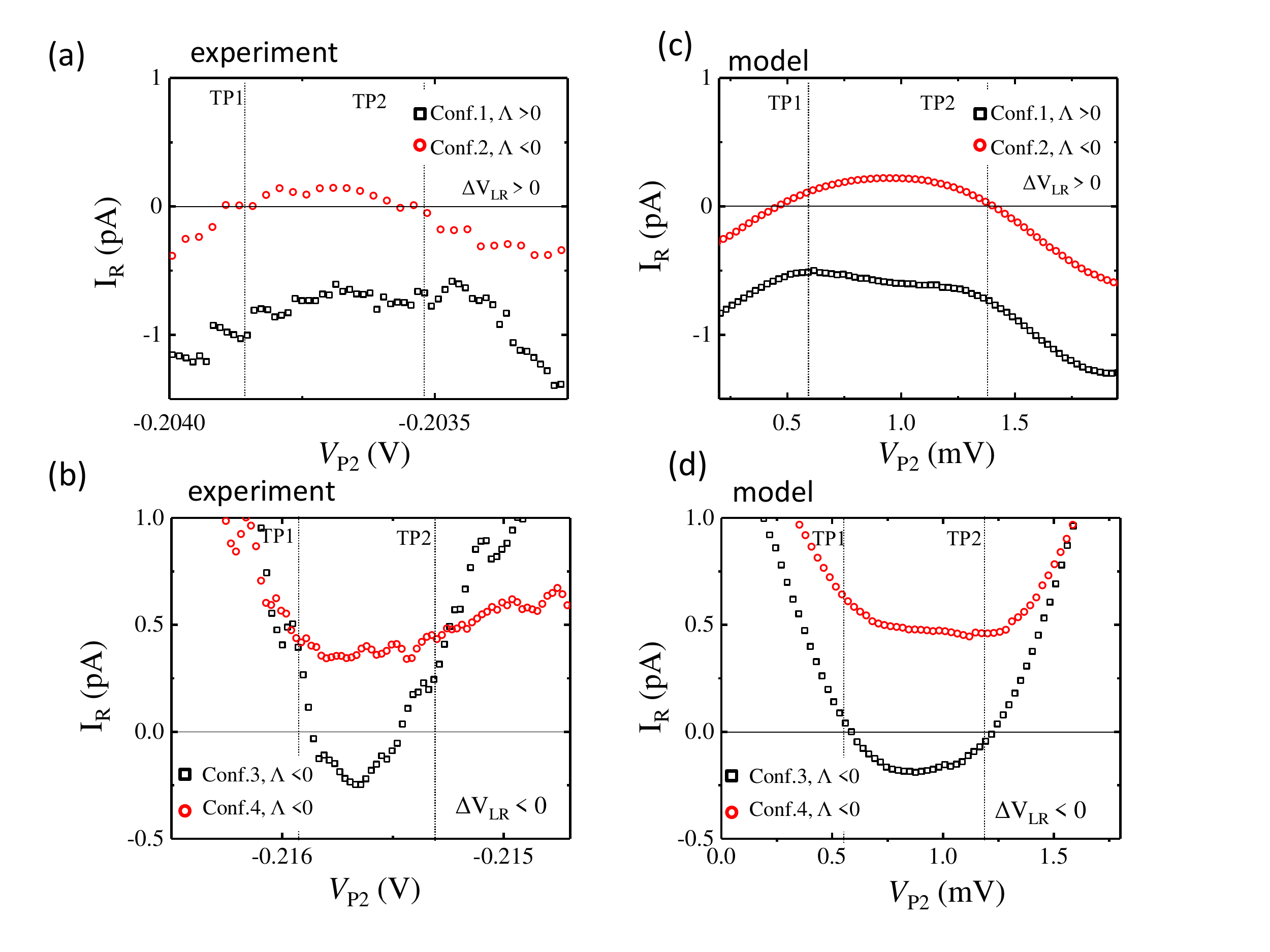}
	\caption{Charge currents in reservoir R due to heating of reservoir H by $\Delta T \approx 100$ mK along the axis of total energy of a stability vertex. The data are plotted against VP2. The positions of the triple points TP1 and TP2 are indicated. Black squares (Red circles) denote configurations for which $\Lambda > 0$ ($\Lambda < 0$). (a) and (b) experiments, (c) and (d) model calculations. Note that for (a) and (c) $\Delta V_{\rm LR} > 0$ while for (b) and (d) $\Delta V_{\rm LR} < 0 $.}
	\label{HCC}
\end{figure}

Equation \eqref{HCC current} indicates that $I$ becomes maximal if $\Lambda$ becomes unity, i.e. if, for example, $\Gamma^{(1)}_L \Gamma^{(0)}_R \gg \Gamma^{(1)}_R \Gamma^{(0)}_L$.
In this case, every energy package which is extracted from reservoir H is used to transfer one electron from R to L. Thus, the QD systems operates as a heat engine with optimal efficiency.

However, useful work is only performed when the heat engine operates against a load, e.g. an external resistor or, equivalently, a potential difference $\Delta V_{\rm LR}$. The engine's efficiency $\eta$ is then given by the ratio of the generated electrical power and the heat flow, $\eta = (I \Delta V_{\rm LR})/J_\rmH$. A straight forward derivation given in Ref.\cite{Sanchez_Optimal_2011} yields that the maximum load is given by a critical voltage $\Delta V_{\rm LR}=V_S$ at which the engine reaches a stable state. In this configuration the rates are such that voltage driven and heat driven charge current cancel each other. Remarkably, for the ideal case $\Lambda = \pm 1$ this is the configuration for which the QD heat engine operates at Carnot efficiency $\eta_{\rm C}=1-T/T_ {\rm H}$. This can be understood intuitively if one considers an analogy to the classical thermodynamic Carnot engine: The Carnot cycle \cite{Fermi_Thermodynamics_1956}, which is also a 4-step-process, consists of two isothermals, during which the systems exchanges heat with the heat baths (similar to steps 2 and 4 in our QD heat engine), and two adiabatics during which (by definition) no heat is transferred but \textit{only} mechanical work is performed on the system or by the system (corresponding to step 1 and 3). For the Carnot cycle it is crucial that those are true adiabatics because only then vanishes the total entropy production of the cycle and $\eta_{\rm C}$ is reached. Similarly, the QD engine operates at $\eta_{\rm C}$ if no entropy is produced during steps 1 and 3. However, this is only the case if the voltage load equals the energy difference $E_{\rm C}$ so that electrons in the conductor system are removed and injected only at the electro-chemical potentials of the reservoirs L and R.
For the non-ideal case $|\Lambda| < 1$, $\eta$ is smaller because processes which counteract current generation, for example by producing heat or transferring charge in the opposite direction, occur with a finite probability~\cite{Sanchez_Optimal_2011}. 

The functionality of a real QD energy harvester has recently been demonstrated \cite{Thierschmann_ThreeTerminal_2015} on a system similar to the one shown in Fig.\ref{Fig1}. In these experiments different barrier configurations had been prepared for the conductor system. Configuration 1 is obtained by increasing the potential barrier connecting QD2 and R by applying a more negative potential to the corresponding gate (gate $G_{\rm R}$ in Fig.~\ref{Fig1}). This strongly reduces the associated tunnelling coefficients compared to those of the barrier connecting QD2 to L and thus ensures broken left-right symmetry. In order to control the energy dependence of the barriers, an asymmetric potential landscape around QD2 is created using gate electrodes in the vicinity of the barrier, for example gate S in Fig.~\ref{Fig1}(c). Next, configuration 2 is prepared. This configuration exhibits the opposite left-right asymmetry compared to configuration 1: reservoir L is pinched off more strongly from QD2 than reservoir R, thus inverting $\Lambda$. Furthermore, two configuration 3 and 4 are adjusted, for which the barriers exhibited different energy dependences.

In order to observe energy harvesting the region of interest is at the centre of the thermal gating clover-leaf pattern. Here the QD energies are aligned in such a way that heat flow between the systems is expected to be largest. In contrast to, for example, thermal gating the potential difference $\Delta V_{\rm LR}$ is to be kept small in order to obtain a large output current. 

Fig~\ref{HCC} (a) and (b) depict the charge current measured in reservoir R upon heating reservoir H by $\Delta T = 100 {\rm mK}$. The data are obtained along the \textit{axis of total energy}, as described in Fig.~\ref{ThermalGating_1}(b), for the different configurations. In all cases a finite signal due to thermal gating is visible outside the TPs (vertical dotted lines) which results from a small $\Delta V_{\rm LR}$. However, now the focus lies on the region between the TPs. 
Figure~\ref{HCC} (a) shows the current measured for configuration 1 (black squares). Here a negative current is observed between TP1 and TP2. Note that currents should be zero if only thermal gating was active. As shown in Ref.\cite{Thierschmann_ThreeTerminal_2015} the current between the TPs maintains its sign upon reversal of $\Delta V_{\rm LR}$. However, it becomes positive if the conductor system is tuned to configurations 2 (red circles). This strongly suggests that the observed current is indeed generated according to the proposed mechanism of heat conversion. Moreover, this nicely shows the validity of Eq.~\ref{HCC current} because both configurations 1 and 2 exhibit tunneling asymmetries with opposite sign. Note that the sign of thermal gating is not affected by the change of barrier asymmetry. 
As shown in Fig.~\ref{HCC} (b) a similar effect is observed for changing the energy dependence of the barriers by variation of gate S (configurations 3 and 4).

The experimental results are well described with the model introduced in section 2. Fig.~\ref{HCC} (c) displays calculations using parameters extracted from the experiments shown in (a), such as tunneling coefficients at the Fermi level, $\Delta T$, $T$, $\Delta V_{\rm LR}$ and $E_{\rm C}$. Fig. (d) gives the calculations corresponding to the experiments shown in (b). 
In both cases, adding a small energy dependences to the barriers produces nice agreement between model and experiments.
 
The applied bias $\Delta V_{\rm LR}$ was small in the experiments ($|\Delta V_{\rm LR}| < 10~ \mu V$) \cite{Thierschmann_ThreeTerminal_2015}. Nevertheless, it constitutes a load against which the energy harvester has to perform work. Thus, $\Delta V_{\rm LR}$ can be used to estimate the efficiency $\eta$ of the energy harvester. For the experiments in Ref.\cite{Thierschmann_ThreeTerminal_2015} $\eta$ was reported to be only small fraction of $\eta_C$. This is related to the fact that the tunnelling asymmetries were relatively small which allowed a large degree of spurious tunneling processes which do not contribute to current generation. Clearly, this points out the route towards device optimization. We expect that already a small enhancement of $\Lambda$ will lead to a significant increase of $\eta$. 

A novel feature of the QD energy harvester is that the direction of its output current is not strictly determined by the temperature gradient. This is a direct consequence of the three-terminal geometry which allows the charge current to be generated between two cold reservoirs. Hence, it can be freely adjusted by choosing the desired orientation of reservoirs L and R during device design. Note that this is fundamentally different to any conventional Seebeck-based thermoelectric device where J and I are intrinsically coupled because they are both carried by the same particles. In the QD energy harvester this intimate coupling is broken up. This might point out a route to overcome the inherent problem of parasitic heat flow in today's thermoelectric devices and thus could give a new momentum for developing the next generation of thermoelectric devices.

Moreover, a similar concept of asymmetric transmission coefficients has recently been utilized to rectify fluctuations in mesoscopic cavities \cite{Roche_Harvesting_2015}. Furthermore, experiments indicate that in QD systems voltage noise \cite{Hartmann_Voltage_2015} and current noise \cite{Bischoff_Measurements_2015} might become rectified. This shows the rich potential for energy harvesting that lies within multi-terminal devices with built in or even tunable asymmetries at the mesoscopic scale.

\section{Summary}
To summarize, we have reviewed recent experiments on thermoelectrics with Coulomb coupled Quantum dots in a three terminal geometry together with a theoretical description.
The experiments have been performed on split-gate defined devices on GaAs/AlGaAs heterostructures at cryogenic temperatures. The system was designed such that one of the terminals serves as a heat bath whose temperature is controlled by means of current heating. The other two reservoirs stay at base temperature and form the conductor circuit. Coupling between the terminals takes place only via Coulomb coupling of the quantum dots. 

In such a system charge flow in the conductor can be controlled by variation of the temperature in the heat reservoir. This effect is called thermal gating \cite{Thierschmann_Thermal_2015}. It was found to be related to enhanced occupation fluctuations on the hot QD which translate into a change of transport properties of the conductor dot. Theoretical considerations indicate that this effect may also be applicable to heat currents in the conductor thus pointing out a route towards an all-thermal transistor.

Moreover, the system has been demonstrated to operate as a quantum dot based energy harvester\cite{Thierschmann_ThreeTerminal_2015}. For these experiments the ability to tune the properties of the structure via gate voltages has turned out to be advantageous because the performance of the device strongly depends on the asymmetry of tunneling coefficients in the conductor. Due to its flexibility in tuning system parameters, it has been possible to modify and even invert the direction of the generated current by adjusting the tunneling asymmetry via external gate voltages. This provides evidence for a decoupling of heat and charge flow in the device. 
The advances reviewed in this article emphasize the potential that lies within multi-terminal thermoelectrics. They demonstrate that new thermoelectric effects can arise in such structures which are fundamentally different from Seebeck-based thermoelectrics and thus may point out a route to overcome limitations inherent to conventional energy harvesters.

\section{Acknowledgements}
We thank F. Arnold, M. Mitterm\"uller and L. Maier for help with device fabrication and measurements and C. Heyn and W. Hansen for wafer growth. This research has been funded by the Deutsche Forschungsgemeinschaft SPP 1386. R. S\'anchez acknowledges support from the Spanish MICINN Grant No. MAT2014-58241-P, COST Action MP1209 ``Thermodynamics in the quantum regime''.

\end{document}